\documentclass[aps,prl,twocolumn]{revtex4}
\usepackage{amssymb}
\usepackage[T2A]{fontenc}
\usepackage[cp1251]{inputenc}
\usepackage{amsmath}
\usepackage[english]{babel}
\usepackage{graphicx}

\input{tcilatex}

\begin{document}

\title{The nonlinear damping of Bose-Einstein condensate oscillations at
ultra-low temperatures}
\author{Yu. Kagan and L.A. Maksimov}
\affiliation{``RRC Kurchatov Institute'', 123182 Moscow, Russia}

\begin{abstract}
We analyze the damping of the transverse breathing mode in an elongated trap
at ultralow temperatures. The damping occurs due to the parametric resonance
entailing the energy transfer to the longitudinal degrees of freedom. It is
found that the nonlinear coupling between the transverse and discrete
longitudinal modes can result in an anomalous behavior of the damping as a
function of time with the partially reversed pumping of the breathing mode.
The picture revealed explains the results observed in \cite{ch}.
\end{abstract}

\maketitle

The problem of the damping of oscillations in a Bose-condensate at $T=0$
and, at least, at ultra-low temperatures within $T<\mu \ll T_{c}$, $\mu $
being chemical potential and $T_{c}$ being Bose-condensation temperature,
represents one of most interesting problems in the physics of Bose-Einstein
condensation. The specific feature of the problem is associated with an
isolation of the system from the environment and, therefore, absence of any
sort of thermal bath. In this aspect the problem has much in common with the
problem of the loss of coherence and the origin of energy dissipation in an
isolated quantum many-particle system. Vast majority of experimental \cite%
{jin}-\cite{mar} and theoretical \cite{pt-st}-\cite{wi} efforts in this
field are referred to the range of relatively high temperatures $T\gg \hbar
\omega _{i}$ where $\omega _{i}$ are the frequencies of a parabolic trap.
Under these conditions normal excitations in essence form an interior
thermostat. The damping in the case concerned is associated with the
interaction of oscillations and thermal excitations accompanying by direct
transfer of energy to the thermal cloud.

In the work of authors \cite{Yu} the problem is considered for the damping
of radial oscillations of the condensate in an elongated trap at $T=0$. The
choice of the trap geometry was not casual. As is found in \cite{kss} for a
2D isotropic harmonic trap, the non-linear Schr\"{o}dinger
(Gross-Pitaevskii) equation has a scaling-invariant solution for the
evolution of the condensate and non-condensate subsystems at an arbitrary
variation of $\omega (t)$ in time. In this case the breathing mode in the 2D
system does not decay irrespective of the magnitude of the amplitude. It was
shown \cite{pr} that the unique properties result from the presence of a
hidden symmetry in the description of such 2D system.

On the other hand, it is found in \cite{Yu} that the transverse breathing
mode can experience a specific damping in an elongated trap. The origin of
such damping lies in the parametric resonance connecting the breathing mode
with longitudinal sound excitations. The parametric resonance arises from
the radial oscillations of the density, resulting in oscillation of the
longitudinal sound velocity. In fact, the results obtained give an
appropriate description only for the initial stage of damping, see below.

Recently, it has been published very interesting experimental investigation
on the damping of the breathing mode in an elongated trap at ultra-low
temperatures $T<\mu \ll T_{c}$ \cite{ch}. The work contains a number of
remarkable results. For the limited initial amplitude of oscillations
characterized by the scale of relative variation $\left| \delta R_{\bot
}/R_{\bot }\right| \approx 0.03-0.04$ for the transverse Thomas-Fermi
radius, the authors have observed the slow and monotonous decay with the
record quality factor $Q=\omega _{0}/\gamma _{0}\geq 2000$. Here $\omega
_{0}=2\omega _{\bot }$ and $\gamma _{0}$ are the frequency of breathing mode
and the damping rate, respectively. The temporal scale of damping is about 1
s. However, provided the initial amplitude is increased by a factor of
three, the whole picture changes drastically. Here on the temporal scale of
200 ms the oscillation amplitude reduces by about three times. Next, the
amplitude unexpectedly starts to grow, reaching the maximum of about
one-half of the initial value, and then decays. The aim of the present study
is to explain an anomalous picture of damping the breathing mode observed in 
\cite{ch}.

The standard theory of parametric resonance implies an invariability for the
amplitude of oscillation of one of parameters in the system. In the general
case this corresponds to the steady feeding of energy. In the isolated
system the situation is different. The amplitude of breathing mode $b_{\perp
}$ decreases in the course of the energy transfer into the subsystem of
longitudinal excitations due to parametric resonance. In these conditions a
nonlinear coupling between the transverse and longitudinal modes plays
important role. If the parametric growth of the amplitude of longitudinal
modes advances their relaxation, after a noticeable reduction of $b_{\bot }$
one can expect a reverse nonlinear pumping of the transverse mode. As a
result, the growth of $b_{\bot }$ starts and only later it will be succeeded
again by the normal damping of the breathing mode. A specific feature of the
parametric resonance is its dependence on a product $b_{\bot }b_{\parallel }$
of the transverse and longitudinal modes. For the smaller magnitude of $%
b_{\bot }\left( t=0\right) $, if a typical time for the pumping of the
longitudinal mode proves to be comparable with the relaxation time or
larger, the damping regime changes significantly and the amplitude of the
breathing mode decays gradually in time. The picture described for the
nonlinear damping allows to explain the results observed in \cite{ch}.

For the quantitative description of the picture of nonlinear damping, we
consider a simple model assuming that the parametric amplification in an
elongated trap is experienced by a single longitudinal mode with frequency $%
\Omega_{\parallel}$ close to a half of the transverse mode frequency $%
\Omega_{\bot}$ and inverse relaxation time $\gamma$. We assume that the
system is in the Thomas-Fermi regime and only long wavelength longitudinal
sound modes are involved into the process.

The Hamiltonian of interacting excitations in the Bose-Einstein condensate
within second quantization for the representation of noninteracting phonons
can be written as (see, \cite{lp}) 
\begin{equation}
\begin{array}{c}
\hat{H}=\hat{H}_{0}+\hat{H}^{\prime }, \\ 
\hat{H}_{0}=\sum\limits_{s}\hbar \Omega _{s}\left( \hat{c}_{s}^{+}\hat{c}%
_{s}+\frac{1}{2}\right) ,\;\;\;\hat{H}^{\prime }=\frac{1}{2}m\int d^{3}x\hat{%
\vec{V}}\hat{n}^{\prime }\hat{\vec{V}}.%
\end{array}
\label{1}
\end{equation}%
The operators of the alternating fraction of density $\hat{n}^{\prime }$ and
velocity $\hat{\vec{V}}$ have the form%
\begin{eqnarray}
\hat{n}^{\prime } &=&\sum\limits_{s}i\left( \frac{\hbar \Omega _{s}n_{00}}{%
2mc_{0}^{2}}\right) ^{1/2}\left( \hat{c}_{s}\chi _{s}-\hat{c}_{s}^{+}\chi
_{s}^{\ast }\right) ,  \label{2} \\
\hat{\vec{V}} &=&\sum\limits_{s}\left( \frac{\hbar c_{0}^{2}}{2m\Omega
_{s}n_{00}}\right) ^{1/2}\left( \hat{c}_{s}\nabla \chi _{s}+\hat{c}%
_{s}^{+}\nabla \chi _{s}^{\ast }\right) .  \label{3}
\end{eqnarray}

In these expressions $n_{00}$ is the magnitude of density $n_{0}\left( \vec{x%
}\right) $ at $\vec{x}=0$, $c_{0}^{2}=\mu /m$, and $m$ is the atom mass. The
real eigenfunctions $\chi _{s}\left( \vec{x}\right) $ for noninteracting
phonons in a trap are solutions of the equation (cp.\cite{st96})%
\begin{equation}
\Omega _{s}^{2}\chi _{s}(\vec{x})+c_{0}^{2}\nabla \left( f\left( \vec{x}%
\right) \nabla \chi _{s}\left( \vec{x}\right) \right) =0,  \label{4}
\end{equation}%
where $f\left( \vec{x}\right) =n_{0}\left( \vec{x}\right) /n_{00}$.

Let us choose the breathing mode and one longitudinal mode in the excitation
spectrum and, correspondingly, eigenfunctions $\chi _{\bot }(\vec{x})$ and $%
\chi _{\parallel }(\vec{x})$. Accordingly, in this case there are two types
of the vertices with integrals like $J=\int d^{3}x\chi _{\bot }(\vec{x}%
)\nabla \chi _{\parallel }(\vec{x})\nabla \chi _{\parallel }(\vec{x})$ and $%
J^{\prime }=\int d^{3}x\chi _{\parallel }(\vec{x})\nabla \chi _{\parallel }(%
\vec{x})\nabla \chi _{\bot }(\vec{x})$ In the Thomas-Fermi approximation the
function $f(x)$ in Eq. (\ref{4}) equals%
\begin{equation}
f(\vec{x})=1-\frac{r^{2}}{R_{\perp }^{2}}-\frac{z^{2}}{R_{\Vert }^{2}},
\label{5}
\end{equation}%
where $R_{\bot }=\sqrt{2\mu /m\omega _{\perp }^{2}}$ and $R_{\Vert }=\sqrt{%
2\mu /m\omega _{\Vert }^{2}}$ are the transverse and longitudinal
Thomas-Fermi radii.

Let wave functions $\chi _{s}$ be normalized by the Thomas-Fermi volume $V=%
\frac{4}{3}\pi R_{\perp }^{2}R_{\Vert }$. Then, for function $\chi _{\bot }$
we find readily from Eq.(\ref{4}),%
\begin{equation}
\chi _{\perp }=\sqrt{\frac{35}{8V}}\left( 1-2\frac{r^{2}}{R_{\perp }^{2}}-%
\frac{z^{2}}{R_{\Vert }^{2}}\right) .  \label{7}
\end{equation}%
In first approximation in the limit $\omega _{\perp }^{2}\gg \omega _{\Vert
}^{2}$ function $\chi _{\parallel }$ depends on $z$ alone (e.g.,\cite{st}).
Hence, calculating integral $J$\ with use of (\ref{7}), we arrive at $\left|
J\right| \sim \omega _{\perp }^{2}/\sqrt{V}c_{0}^{2}$. On the other hand,
the estimate of $J^{\prime }$ yields $\left| J^{\prime }\right| \sim \omega
_{\parallel }^{2}/\sqrt{V}c_{0}^{2}$. In the strongly elongated trap we can
retain only the terms with $J$ in $H^{\prime }$. As a result, in Hamiltonian 
$H^{\prime }$ there are only two terms from eight ones that can realize a
quasi resonance coupling between these selected modes. Retaining only these
terms, we represent the expression for $H^{\prime }$ in the explicit form%
\begin{eqnarray}
\hat{H} &=&i\hbar gJ(\hat{c}_{_{\Vert }}^{+}\hat{c}_{_{\Vert }}^{+}\hat{c}%
_{\perp }-\hat{c}_{\perp }^{+}\hat{c}_{_{\Vert }}\hat{c}_{_{\Vert }}),
\label{8} \\
g &=&\frac{1}{4}\left( \frac{\hbar c_{0}^{2}\Omega _{\perp }}{2mn_{00}\Omega
_{\Vert }^{2}}\right) ^{1/2}.  \label{9}
\end{eqnarray}%
In the Heisenberg representation%
\begin{equation*}
\frac{\partial }{\partial t}\hat{c}_{s}=-\frac{i}{\hbar }[\hat{c}_{s},\hat{H}%
].
\end{equation*}%
Using Eqs. (\ref{1}) and (\ref{8}), we find%
\begin{equation}
\begin{array}{c}
\frac{\partial }{\partial t}\hat{c}_{\Vert }=-i\Omega _{\Vert }\hat{c}%
_{\Vert }+2gJ\hat{c}_{\Vert }^{+}\hat{c}_{_{\perp }} \\ 
\frac{\partial }{\partial t}\hat{c}_{\perp }=-i\Omega _{\perp }\hat{c}%
_{\perp }-gJ\hat{c}_{_{\Vert }}\hat{c}_{_{\Vert }}%
\end{array}
\label{10}
\end{equation}

Let us go over into the classical Bose field in equations (\ref{10}). This
is a conventional transformation in treating excitations in the
Bose-Einstein condensate, see, e.g., \cite{da}. In addition, we note that
even for the extremely low temperature 40 nK \cite{ch}, there are 5 phonons
at the longitudinal mode and hundreds at the excited transverse mode. Within
the framework of the standard procedure 
\begin{equation}
\hat{c}_{\Vert }\rightarrow b_{\parallel }e^{-i(\Omega _{\Vert }t-\varphi
_{\Vert })},\hat{c}_{\perp }\rightarrow b_{\perp }e^{-i(\Omega _{\perp
}t-\varphi _{\perp })}.  \label{120}
\end{equation}%
Here $b_{s}$ and $\varphi _{s}$ are the slowly varying real quantities.
Performing this transformation and separating the real and imaginary parts,
we find%
\begin{eqnarray}
\frac{\partial }{\partial t}\bar{b}_{\parallel } &=&-\alpha \bar{b}_{\bot }%
\bar{b}_{\parallel }\cos \varphi -\gamma (\bar{b}_{\parallel }-\bar{b}%
_{\parallel }(0)),  \label{181} \\
\frac{\partial }{\partial t}\bar{b}_{\bot } &=&\frac{1}{2}\alpha \bar{b}%
_{\parallel }^{2}\cos \varphi ,  \label{182} \\
\frac{\partial }{\partial t}\varphi &=&\Delta \Omega +2\bar{b}_{\bot }(1-%
\frac{\bar{b}_{\parallel }^{2}}{4\bar{b}_{\bot }^{2}})\sin \varphi .
\label{183}
\end{eqnarray}

Here we have singled out the initial value $b_{\bot }(0)$ for breathing mode
amplitude \ and introduced a relative amplitudes $\bar{b}_{s}=b_{s}/b_{\bot
}(0)$ and the notation%
\begin{equation}
\alpha =2g\left| J\right| b_{\bot }(0),\ \ \ (J<0).  \label{17}
\end{equation}%
The phase $\varphi $ is defined as 
\begin{equation}
\varphi (t)=\varphi _{\perp }(t)-2\varphi _{\Vert }(t)+\Delta \Omega t,\text{%
\ }\Delta \Omega =\Omega _{\perp }-2\Omega _{\Vert .}  \label{171}
\end{equation}%
In the equation for the longitudinal (\ref{181}) component we have
introduced the term taking into account effective relaxation. Quantity $%
b_{\parallel }(0)$ is the initial equilibrium value $b_{\parallel }.$

A set of Eqs. (\ref{181})-(\ref{183}) describes nonlinear damping of the
breathing mode in the general case. Let us consider first the initial period
provided $b_{\parallel }(0)\ll b_{\bot }(0).$ Omitting the ratio $%
(b_{\parallel }/2b_{\bot })^{2}$ and neglecting variation of $b_{\bot }$ in
the Eq. (\ref{183}) we arrive to the equation with separable variables. The
solution of this equation can readily be found. For $\left| \Delta \Omega
\right| <2\alpha $%
\begin{equation}
\ln \frac{\Delta \Omega \tan \frac{\varphi }{2}+2\alpha -\varkappa }{\Delta
\Omega \tan \frac{\varphi }{2}+2\alpha +\varkappa }=\varkappa t+C,
\label{16}
\end{equation}%
where $\varkappa \ =\sqrt{\left( 2\alpha \right) ^{2}-\left( \Delta \Omega
\right) ^{2}}$. For $\varkappa t\gg 1$, phase $\varphi $ tends to a constant
value. For $\Delta \Omega \rightarrow 0$, $\varphi \rightarrow \pi $ and in
Eqs. (\ref{181}), (\ref{182}) $\cos \varphi \rightarrow -1$. As a result,
solution of Eq. (\ref{181}) can be represented as $b_{\parallel }\propto
\exp [\left( \alpha -\gamma \right) t]$. An exponential growth with smaller
decrement ($\left| \cos \varphi \right| <1)$ holds for finite $\Delta \Omega 
$ if $\ \left| \Delta \Omega \right| <2\alpha $ and $\alpha \left| \cos
\varphi \right| >\gamma $. This is a typical parametric resonance resulting
in the damping of the transverse oscillations in the condensate \cite{Yu}.

For an arbitrary time, the analysis requires a combined solution for all
three nonlinear equations (\ref{181})-(\ref{183}).

Let us find relation between the coefficients in this system and the
physical parameters. The vibrational energy, associated with the breathing
mode, equals $E_{vib}\approx \frac{4}{3}\mu N(\delta R_{\perp }/R_{\perp
})^{2},$ where $N$ is the total number of particles. On the other hand, in
accordance with definition (\ref{120}), one has $E_{vib}=\hbar \Omega _{\bot
}b_{\bot }^{2}$. Thus $b_{\bot }(0)=\left( 4\mu N/3\hbar \Omega _{\bot
}\right) ^{1/2}\left\vert \delta R_{\perp }/R_{\perp }\right\vert _{0},$
where $\left\vert \delta R_{\bot }/R_{\bot }\right\vert _{0}$ is an initial
value of a relative amplitude of the transverse oscillations. Estimating
integral $J$ at the vertex of the interaction Hamiltonian (\ref{8}), we find 
$J\approx -0.4\Omega _{\bot }^{2}/V^{1/2}c_{0}^{2}.$

Hence, determining coefficient $\alpha $ (\ref{17}) and involving value of $%
g $ (\ref{9}), we have 
\begin{equation}
\alpha \approx 0.3\omega _{\bot }\left\vert \frac{\delta R_{\perp }}{%
R_{\perp }}\right\vert _{0}  \label{19}
\end{equation}%
In derivation we used the following relations $N=\left( 2/5\right) n_{00}V$, 
$\Omega _{\parallel }\approx \Omega _{\bot }/2=\omega _{\bot }$.

The parametric resonance requires nonzero amplitude $b_{\parallel }(0)$. For 
$T=0$, $b_{\parallel }\left( 0\right) $ is the amplitude of zero-point
oscillations. For $T>\hbar \Omega _{\parallel }$ we have $b_{\parallel
}\left( 0\right) \approx (T/\hbar \Omega _{\parallel })^{1/2}$.

Let us turn to the relaxation in the longitudinal subsystem. The familiar
dispersion law for longitudinal excitations $\omega _{k}^{2}=\omega
_{\parallel }^{2}k(k+3)/4$ \cite{st},\cite{fl} in combination with the
mesoscopic character of the system results in a peculiar picture of
relaxation. For a strongly elongated trap $\omega _{\bot }\gg \omega
_{\parallel }$, frequency $\Omega _{\parallel }$ is met with large value $%
k\gg 1$. This state can decay into numerous pairs of excitations with
integer indices satisfying condition $k_{1}+k_{2}=k$. It is interesting that
the mismatch of resonances for different pairs changes only within interval $%
\Delta \omega =\omega _{\parallel }/2\div 3\omega _{\parallel }/4$. Within
the framework of the hydrodynamic approximation these processes are
described by the same Hamiltonian $H$ (\ref{1}) involving (\ref{2}) and (\ref%
{3}). Selecting a part of the interaction Hamiltonian responsible for the
evolution of state $k$, we have 
\begin{equation}
\hat{H}^{\prime }=\hbar \sum_{k^{\prime }}B_{k^{\prime }}(\hat{c}%
_{_{k-k^{\prime }}}^{+}\hat{c}_{_{k^{\prime }}}^{+}\hat{c}_{k}-\hat{c}%
_{k}^{+}\hat{c}_{_{k^{\prime }}}\hat{c}_{_{k-k^{\prime }}})
\end{equation}%
A direct estimate of vertex $B_{k^{\prime }}$ yields%
\begin{equation*}
B_{k^{\prime }}\approx 2i\left( \frac{\hbar }{\mu N}\right) ^{1/2}\left(
\omega _{k}\omega _{k^{\prime }}\omega _{k-k^{\prime }}\right) ^{1/2}.
\end{equation*}%
In the usual conditions $\left| B_{k^{\prime }}(k)\right| \ll \Delta \omega $%
. At $T=0$ the excited $k$-state spreads partially over $k/2$ pairs of
states (for simplicity we imply that $k$ is even) with the inverse time of
reducing the density%
\begin{equation}
\frac{1}{\tau _{k}}\approx \sum_{k^{\prime }}^{k/2}\frac{\left| B_{k^{\prime
}}(k)\right| ^{2}}{\Delta \omega _{k^{\prime }}}.  \label{21}
\end{equation}

All $k$ states appeared spread, in turn, over a large number of states. In
these conditions, taking also into account the scatter of $\Delta \omega
_{k^{\prime }}$ and vertices $B_{k^{\prime }}$, one can suppose that the
probability for revival of $k$-state with a noticeable amplitude is
negligible. In essence, relaxation occurs for the time about $\tau _{k}$ (%
\ref{21}).

As concerns finite temperatures $T>\hbar \omega _{k}$, the role of
transitions with the involvement of low frequency states enhances. Hereupon, 
$\left| B_{k_{1}}(k)\right| ^{2}$ is multiplied approximately by a factor $%
T/\hbar \omega _{k^{\prime }}+T/\hbar \omega _{k-k^{\prime }}=T\omega
_{k}/\left( \hbar \omega _{k^{\prime }}\omega _{k-k^{\prime }}\right) .$ Then%
\begin{equation}
\frac{1}{\tau _{k}}\approx 2k\omega _{k}\frac{T}{\mu }\frac{1}{N}\frac{%
\omega _{k}}{\Delta \omega }.  \label{22}
\end{equation}%
The parameter $\gamma $ in Eq. (\ref{181}) is $\approx (2\tau _{k})^{-1}.$

Let us consider a solution of the system of nonlinear equations (\ref{181})-(%
\ref{183}). We choose the parameters close to that of \cite{ch}: $\omega
_{\bot }=1150$ s$^{-1}$, $\omega _{\parallel }=70$ s$^{-1}$, $N=10^{5}$, $%
\mu =60$ nK. The finiteness of temperature $T=40$ nK really affects only the
magnitude of parameter $\gamma $, see (\ref{22}). For these values of
parameters $k\approx 30$ and coefficient $\gamma $ lie within interval $%
5\div 10$ s$^{-1}$. We take $\gamma =7$ s$^{-1}$ for calculation. For
initial condition $\left| \delta R_{\bot }/R_{\bot }\right| _{0}=0.09$,
coefficient $\alpha $ (\ref{17}) takes value $\alpha \approx 30$ s$^{-1}$.
The spacing between the neighbor levels in the spectrum of longitudinal
excitations is close to $\omega _{\parallel }/2$. Within the two-level model
under consideration it is naturally to suppose that frequency $\omega _{\bot
},$ falling within this interval, is shifted to one of the levels. For
calculating, we put $\Delta \Omega =15$ s$^{-1}$.

In Fig.1 the results of solving the system (\ref{181})-(\ref{183}) are
presented for the given values of $\alpha $, $\gamma $ and $\Delta \Omega $
as solid lines. The time behavior of $\ \left| \delta R_{\bot }/R_{\bot
}\right| =\bar{b}_{\bot }\left| \delta R_{\bot }/R_{\bot }\right| _{0}$ is
plotted in Fig.1a. A drastic reduction of the breathing mode amplitude for
the relatively short interval of time is a direct manifestation of
parametric resonance. This is especially clear from the burst-like growth of
the amplitude for longitudinal oscillations $\tilde{b}_{\parallel }(t)=\bar{b%
}_{\parallel }(t)\left| \delta R_{\bot }/R_{\bot }\right| _{0}$, as is seen
in Fig.1b. At some time moment the phase $\varphi $ reaches value $3\pi /2$ (%
$\pi /2$ for $\Delta \Omega <0$) and the reverse energy transfer from
longitudinal into the transverse mode starts. In place of reduction, a
nontrivial growth of the breathing mode amplitude starts. The growth is over
when the phase reaches value $5\pi /2$ ($-\pi /2$) and the normal trend
recovers. Note that for the chosen magnitude of initial amplitude $\left|
\delta R_{\bot }/R_{\bot }\right| _{0}$ or parameter $\alpha $, entire
qualitative picture of nonlinear damping holds for a wide variation of $%
\Delta \Omega $.

However, the picture of damping changes drastically if the initial amplitude
is noticeably smaller. In Figs. 1a and 1b the same dependences for the case $%
\left| \delta R_{\bot }/R_{\bot }\right| _{0}=0.035$ are shown with the
dashed lines, the other parameters being unchanged. We see a normal
monotonic behavior for the amplitude reduction $\left| \delta R_{\bot
}/R_{\bot }\right| $. As is seen from Fig. 1b, the damping occurs in the
conditions of the relatively small-expressed parametric resonance. Herewith,
phase $\varphi $ changes within the limited boundaries. 
\begin{figure}
\includegraphics{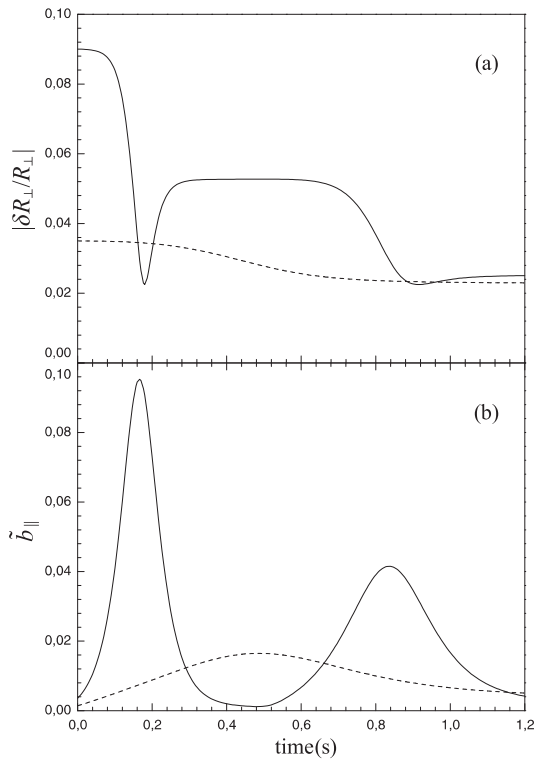}
\caption{\label{fig1} (a) The amplitude of breathing mode $\left\vert \delta R_{\bot }/R_{\bot }\right\vert $ and 
(b) the amplitude of longitudinal mode $\tilde{b}_{\parallel }$ as a function of time for two initial values $\left\vert \delta R_{\bot }/R_{\bot }\right\vert _{0}=0.09$ (---) and $\left\vert \delta R_{\bot }/R_{\bot }\right\vert _{0}=0.035$ (- - -)}
\end{figure}%

The picture of nonlinear damping, found within the framework of solving a
relatively simple model, explains the results observed in \cite{ch}. We have
chosen intentionally the magnitude of parameters close to that of \cite{ch}
in order to make comparison more adequate. At the same a detailed
quantitative comparison requires more elaborate model.

It is worthwhile to note that an additional damping due to extrinsic reasons
can be significant for the slow dynamics at small value of amplitude $%
\left\vert \delta R_{\bot }/R_{\bot }\right\vert _{0}$. However, this
damping will not play any noticeable role in the case of large magnitude of
the amplitude. In this aspect the nonlinear damping is a purely intrinsic
phenomenon.

It seems that the results obtained have a relatively general character. In
fact, the typical feature of all systems used for studying Bose-Einstein
condensation in ultracold gases is their mesoscopic character and separation
from environment. In these conditions the problem of the damping of coherent
oscillations in the condensate at $T=0$ or, at least, $T<\mu \ll T_{c}$
acquires a special sounding. On one hand, this is connected with the problem
of revealing the decay channel. On the other hand, provided such channel
exists, the energy is transferred to the discrete degrees of freedom. In
this case the analysis of energy relaxation requires involving nonlinear
coupling between modes for the both dissipationless and dissipative
kinetics. The problem considered in the present work is an instructive
example of this general problem.

This work is supported by INTAS (project 2001-2344), by Netherlands
Organization for Science Research (NWO), and by the Russian Foundation for
Basic Research.

\end{document}